\documentclass{jfm}

\usepackage{graphicx}
\usepackage{natbib}
\usepackage{epstopdf}
\usepackage{amsmath}
\usepackage{esdiff}
\usepackage{subfig}
\usepackage{float}
\usepackage{latexsym}

\ifCUPmtlplainloaded \else
  \checkfont{eurm10}
  \iffontfound
    \IfFileExists{upmath.sty}
      {\typeout{^^JFound AMS Euler Roman fonts on the system,
                   using the 'upmath' package.^^J}%
       \usepackage{upmath}}
      {\typeout{^^JFound AMS Euler Roman fonts on the system, but you
                   dont seem to have the}%
       \typeout{'upmath' package installed. JFM.cls can take advantage
                 of these fonts,^^Jif you use 'upmath' package.^^J}%
      }
  \else
  \fi
\fi

\ifCUPmtlplainloaded \else
  \checkfont{msam10}
  \iffontfound
    \IfFileExists{amssymb.sty}
      {\typeout{^^JFound AMS Symbol fonts on the system, using the
                'amssymb' package.^^J}%
       \usepackage{amssymb}%

      }{}
  \fi
\fi

\ifCUPmtlplainloaded \else
  \IfFileExists{amsbsy.sty}
    {\typeout{^^JFound the 'amsbsy' package on the system, using it.^^J}%
     \usepackage{amsbsy}}
    {\providecommand\boldsymbol[1]{\mbox{\boldmath $##1$}}}
\fi

\newcommand\Kn{\mbox{\textit{Kn}}}  

\newsavebox{\astrutbox}
\sbox{\astrutbox}{\rule[-5pt]{0pt}{20pt}}

\title{Gaseous viscous peeling of linearly elastic substrates}
\author{Shai B. Elbaz, Hila Jacob  and Amir D. Gat}
\affiliation{Faculty of Mechanical Engineering, Technion - Israel Institute of Technology, Haifa 3200003, Israel}

\pubyear{2017} \volume{999} \pagerange{999--9999}
\date{2017}
\begin{document}
\maketitle


\abstract{We study pressure-driven propagation of gas into a micron-scale gap between two linearly elastic substrates. Applying the lubrication approximation, the flow-field is governed by the interaction between elasticity and viscosity, as well as weak rarefaction and low-Mach-compressibility, characteristic to gaseous microflows. Several physical limits allow simplification of the governing evolution equation and enable solution by self-similarity. These limits correspond to different time-scales and physical regimes which include compressiblity-elasticity-viscosity, compressiblity-viscosity and  elasticity-viscosity dominant balances.  For a prewetting layer thickness which is similar to the elastic deformation generated by the background pressure, a symmetry between compressibility and elasticity allows to obtain a self-similar solution  which includes weak rarefaction effects. The results are validated by numerical solutions of the evolution equation.}

\section{Introduction}
In this work we analyze the propagation of a Newtonian ideal gas into a thin gas-filled gap, with thickness of the order of microns, bounded by linearly elastic substrates. At standard atmospheric conditions, pressure-driven gaseous flows within micron-sized configurations involve significant viscous resistance, yielding 'low-Mach-compressibility' with negligible inertial effects \citep{Saffman.1957,Arkilic.1997}. In addition, weak rarefaction effects emanating from Knudsen numbers at the range of $\Kn\approx 0.01$ to $\approx 0.1$  yield velocity- and temperature-slip at the solid boundaries \citep{Cercignani.Rarefied.Gas.Dynamics}. Thus, gaseous viscous peeling is governed by interaction of elasticity of the boundaries, gas viscosity, low-Mach-compressibility and weak-rarefaction.


The limit of large deformations compared with the initial gap corresponds to viscous peeling dynamics which are characterized by a distinct peeling front, similarly to the fronts in free-surface flows \citep{oron1997long} and gravity currents \citep[e.g.][]{huppert1982propagation}.
\cite{mcewan1966peeling} were the first to examine viscous peeling, and studied the removal of an adhesive strip from a rigid surface. While previous studies modelled the adhesive as a Hookean elastic material, \cite{mcewan1966peeling} examined the opposite limit of a Newtonian viscous fluid, which enabled calculation of the peeling speed as a function of the applied tension. Other works involving viscous peeling dynamics include \cite{hosoi2004peeling}, who examined the peeling and levitation of a elastic sheet over a thin viscous film and \cite{lister2013viscous} who studied axisymmetric viscous peeling of an elastic sheet from a flat rigid surface by injection of fluid between the surface and the sheet \citep[additional relevant works include][]{hodges2002spreading,hewitt2015elastic,thorey2016elastic,elbaz2016axial,young2017long}.

Effects of weak rarefaction and 'low-Mach-compressibility' on pressure driven flows were extensively studied in the context of gaseous micro-fluidics \citep[][]{Gad_el_Hak_mems_review,Ho_and_Tai.1998}. The first experimental works were conducted by \cite{pong1994non} and \cite{liu1995mems} and presented non-constant pressure gradient in uniform micro-channels associated with 'low-Mach-compressibility' effects. \cite{Arkilic.1997} and \cite{Zohar_Straight} analytically and experimentally studied gas flow through a uniform long micro-channel with both compressibiliy and velocity-slip effects  \citep[among others such as][]{Colin.2001,Wereley.2004}.  Gaseous flows through shallow non-uniform micro-channels involving bends, constrictions and cavities were studied experimentally by \cite{Zohar_Cavities,Zohar_Bends,Zohar_constriction} and treated analytically by \cite{gat2008gas,gat2009higher,gat2010gas,gat2010compressible}.

The aim of the current work is to study gaseous viscous peeling dynamics involving low-Mach-compressibility and weak-rarefaction. The structure of this work is as follows: In \S2 we define the problem and develop the evolution equation. In \S3.1 we present an implicit steady-state solution. In \S3.2 we present self-similar solutions of the evolution equation for various limits and map the transitions between the different regimes. In \S3.3 we develop a self-similar solution which includes weak-rarefaction effects for configurations which involve symmetry between elasticity and compressibility. Concluding remarks are presented in \S4.

\section{Problem formulation and derivation of the evolution equation}
\begin{figure}
\centering
\includegraphics[width=0.9\textwidth]{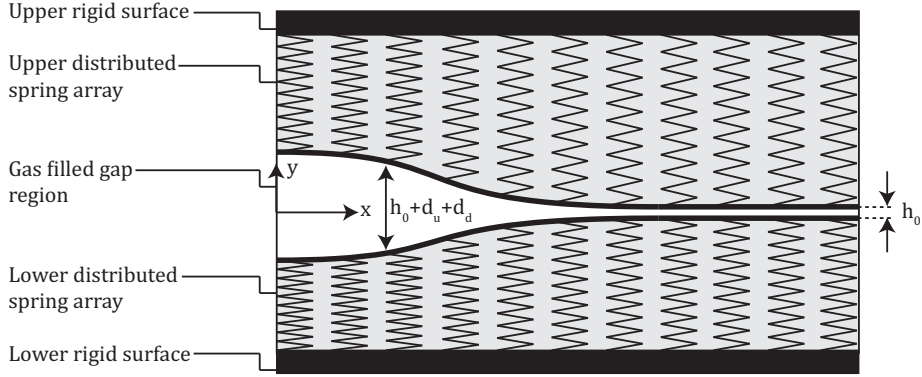}
\caption{A schematic description of the configuration: A gas-filled gap separates two parallel distributed linear spring arrays bounded by rigid surfaces. The initial prewetting gas layer is denoted by $h_0$ and upper and lower vertical deformations are denoted by $d_u$ and $d_d$, respectively.}
\label{figure1}
\end{figure}

We examine pressure-driven gaseous viscous peeling of a two dimensional gap bounded by linear elastic substrates. The configuration (similar to \cite{gaver1996steady}) is illustrated in figure \ref{figure1}. The $x-y$ coordinate system is located at the center of the gap at rest, where $x$ is parallel to the gap streamwise direction, time is $t$, and temperature is $\theta$. At rest, the constant gap between the lower and upper substrates is denoted by $h_0$, and  contains gas at the background pressure $p_a$. Film height is denoted by $h=h_0+d$, where $d=d_u+d_l$ is the combined pressure induced vertical deformation of the upper and lower surfaces. The stiffness coefficients of the upper and lower distributed spring substrates are $k_u$ and $k_l$, respectively, where we define total channel stiffness by $k=(k_u^{-1}+k_l^{-1})^{-1}$. Gas velocity is $(u,v)$, absolute pressure is $p$,  gas viscosity is $\mu$, gas density is $\rho$, the gas constant is $r_g$ and the gas mean-free-path is $\lambda$. 

We define $l$ as the axial length-scale of the configuration and $p^*$ as characteristic  gauge pressure (representing the characteristic value of $p-p_a$). Thus the characteristic elastic displacement is given by $p^*/k$.  We define the dimensionless ratios
\begin{equation}\label{Pis}
Kn=\frac{\lambda}{h_0+d}=Kn_a \left(\frac{p_a}{p}\right)\left(\frac{h_0}{h_0+d}\right),\quad  \Pi_H=\frac{h_0 k}{p^*},\quad  \Pi_P=\frac{p_a}{p^*}, \end{equation}
where $Kn$ is the Knudsen number representing the validity of the continuum assumption;  $\Kn_a$ corresponds to the Knudsen number at the background pressure and the initial gap $h_0$;  $\Pi_H$ is the ratio of initial gap to the elastic displacements and $\Pi_P$ is the ratio of the external background pressure to $p^*$. The limit $\Pi_P\rightarrow\infty$ corresponds to negligible low-Mach-compressibility and the limit $\Pi_H\rightarrow\infty$  corresponds to negligible elastic deformations.

Hereafter we denote normalized variables by Capital letters. Scaling according to the lubrication approximation, the corresponding normalized parameters and variables are the coordinates  $(X,Y)=(x/l,yk/p^*)$, time $T=t (p^*)^2/k^2l^2\mu$, total elastic vertical displacement  $D=d k/p^*$,  film height  $H=h k/p^*=\Pi_H+D$, fluid velocity  $(U,V)=(u k^2\mu l/(p^*)^3,v k^3\mu l^2/(p^*)^4)$, pressure  $P=p/p^*$ and density  $\Lambda=\rho/(p^*/r_g\theta)$. 

We assume isothermal flow and negligible body forces, which is common practice for flows through micron-sized configurations \cite[e.g.][]{Arkilic.2001,Zohar_Straight}. Applying the above scaling, the requirements for the validity of the lubrication approximation are given by the following  relations,
\begin{equation}\label{Small_Relations}
\varepsilon=\frac{p^*}{k l}\ll1,\quad  \alpha^2=\varepsilon Re=\frac{(p^*)^6}{r_g\theta k^4\mu^2 l^2}\ll1
 \end{equation}
where $\varepsilon$ represents the slenderness of the configuration, $\alpha^2$ is the Womersley number and $\varepsilon Re$ is the reduced Reynolds number. Applying (\ref{Small_Relations}) allows to utilize the standard lubrication form of the momentum equations, ${\partial P}/{\partial X} \sim {\partial ^2U}/{\partial {Y^2}}$, ${\partial P}/{\partial Y }\sim 0$,
compressible conservation of mass equation, ${{\partial \Lambda }}/{{\partial T}} + {{\partial \left( {\Lambda U} \right)}}/{{\partial X}} + {{\partial \left( {\Lambda V} \right)}}/{{\partial Y}} = 0
$
and isothermal equation of state $\Lambda=P$.

The validity of the continuum assumption requires a sufficiently small Knudsen number $Kn=\lambda/h_0$, defined here as the ratio between the molecular mean-free-path $\lambda$ and the prewetting layer thickness $h_0$. While for $Kn<10^{-3}$ use of the no-slip boundary condition is appropriate, for gas flows through micron-sized configurations at standard atmospheric conditions $\Kn\approx 10^{-1}-10^{-2}$. This Knudsen regime requires the incorporation of velocity-slip, and thus the boundary conditions are given by the Navier-slip condition,
\begin{subequations} \label{boundary_conditions_sbs}
\begin{equation}
\label{boundary_conditions_b_sbs} \left[U(Y = D_u),U(Y = D_l)\right] =   \sigma \frac{{K{n_a}{\Pi _H}{\Pi_P}}}{P}\left[-\frac{{\partial U(Y= D_u)}}{{\partial Y}},\frac{{\partial U(Y=D_l)}}{{\partial Y}}\right]
\end{equation}
as well as the kinematic boundary condition at the gas-substrate interface
\begin{equation}
\label{boundary_conditions_a_sbs} 
\left[V(Y = D_u),V(Y = D_l)\right] = \frac{1}{2}\left[ {\frac{{\partial D_u}}{{\partial T}} + U(Y = D_u)\frac{{\partial D_u}}{{\partial X}}},{\frac{{\partial D_l}}{{\partial T}} + U(Y = D_l)\frac{{\partial D_l}}{{\partial X}}} \right],
\end{equation}
\end{subequations}
where the coefficient $\sigma$ represents the interaction between the gas molecules and the solid wall \citep{chapman1952mathematical}.

Applying (\ref{boundary_conditions_b_sbs}) to the $X$-momentum equation yields the velocity profile,
\begin{equation}
\label{vel_prof_sbs} 
U=\frac{1}{2}\frac{{\partial P}}{{\partial X}}\left( {{Y^2} - \frac{{{(D+\Pi_H)^2}}}{4} - \frac{{\sigma K{n_a}{\Pi _H}{\Pi_P}}}{2}\frac{(D+\Pi_H)}{P}} \right).
\end{equation}
Integration of mass conservation equation in conjunction with (\ref{boundary_conditions_a_sbs})- (\ref{vel_prof_sbs}), and applying the normalized linear elastic relation $D=P-\Pi_P$,  yields the evolution equation
\begin{multline} \label{GE_sbs_H}
\frac{{\partial }}{{\partial \mathfrak{T}}}\left[(D+\Pi_H)(D + \Pi_P)\right] = \\ \frac{\partial }{{\partial X}}\left[ {\left( {(D + {\Pi _P}){{(D + {\Pi _H})}^3} + 6\sigma K{n_a}{\Pi _H}{\Pi _P}{{(D + {\Pi _H})}^2}} \right)\frac{{\partial D}}{{\partial X}}} \right],
\end{multline}
where hereafter $\mathfrak{T}=T/12$ for convenience. 

\section{Solutions of the evolution equation}
\subsection{Steady-state}
An implicit solution of (\ref{GE_sbs_H}) may be obtained for steady-state flow in a finite configuration with prescribed pressures at the inlet and outlet sections ($D(0)=P(0)-\Pi_P$ and $D(1)=P(1)-\Pi_P$, respectively),
\begin{subequations}\label{steady_sol_sbs_imp1}
\begin{equation}
X(D) = \frac{{F(D) - F(D(0))}}{{F(D(1)) - F(D(0))}},
\end{equation}
where
\begin{equation}
F(D) = 40\sigma {K{n_a}}{\Pi _H}{\Pi_P} {(D+\Pi_H)^3} + 5(D-\Pi_P){(D+\Pi_H )^4} - {(D+\Pi_H )^5}.
\end{equation}
\end{subequations}

\begin{figure}
\centering
\includegraphics[width=0.9\textwidth]{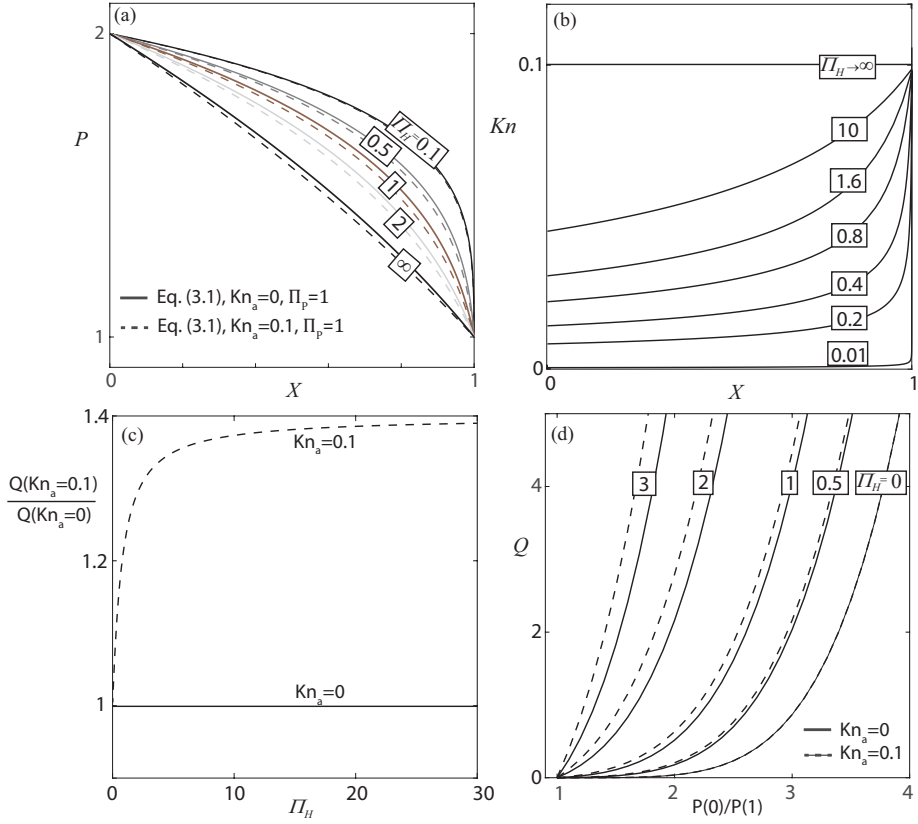}
\caption{Steady-state solution (\ref{steady_sol_sbs_imp1}) of gaseous viscous flow in a 2D gap bounded by linearly elastic substrates. Gas pressure (a) and local Knudsen (b) profiles vs. $X$ for varying $\Pi_H$ for $P(0)/P(1)=2$. (c) Mass flow rate vs. $\Pi_H$ for $Kn_a=0$ and $Kn_a=0.1$. (d) Mass flow rate vs. pressure ratio $P(X=0)/P(X=1)$ for various values of $\Pi_H$. For all panels $Kn_a=0.1$, $\sigma=1$ and $\Pi_P=1$.}
\label{figure2}
\end{figure}

Solution (\ref{steady_sol_sbs_imp1}) is depicted in figure \ref{figure2}(a) for the case of $\Pi_P=1$,  $P(0)/P(1)=2$, $Kn_a=0$ (smooth lines) and  $Kn_a=0.1$ (dashed lines) for various values of $\Pi_H$. Small values of $\Pi_H$ represent large ratios of elastic deformation to initial gap $h_0$, which reduce the pressure gradient near the inlet while increasing it towards the outlet. For constant prewetting layer thickness and background pressure (i.e. constant $Kn_a$), decreasing $\Pi_H$ decreases the local Knudsen number (as seen in panel (b)) and thus decrease the effect of weak rarefaction on the pressure distribution. Panel (c) presents the effect of weak rarefaction on the mass-flow-rate vs. $\Pi_H$. Weak rarefaction effects on mass-flow-rate tend to a constant finite value for $\Pi_H\rightarrow\infty$, and decreases with $\Pi_H$. Panel (d) presents the mass-flow-rate vs. the pressure difference for various values of $\Pi_H$. While the gradient of the different lines vary significantly with $\Pi_H$ for the limit of small pressures at the inlet, as the inlet pressure increases the viscous resistance no longer depends on $\Pi_H$ (or $h_0$) and the gradients converge. 

\subsection{Self-similar Barenblatt solutions for negligible rarefaction effects}
While exact solutions of (\ref{GE_sbs_H}) are not available, several limits involving negligible rarefaction effects yield known self-similar solutions. Furthermore, the flow-field may be described by  different approximate solutions during different time-scales of observation of the peeling process. Physical insight may thus be gained by mapping the different approximate solutions and the corresponding time-scales and transitions.

We focus on fundamental solutions for the case of impulse driven peeling - an abrupt release of a finite mass at $T=0$ into the inlet at $X=0$. This process is characterized by a compactly supported region of displacement and a distinct front, denoted by $X_F$. The relevant integral form of mass conservation is
\begin{equation}\label{impulse_conditions}
\int_0^{{X_F}} {\left[(D+\Pi_H)(D+\Pi_P)-\Pi_H\Pi_P\right]dX = M}\,,
\end{equation}
where $M$ is a constant representing the mass injected into the interface at $T=0$. The conditions near the contact line $X\to X_F$ are $D\to 0$ and $P\to \Pi_P$.


Due to the sudden injection of mass at the inlet, for all values of $\Pi_P,\Pi_H$, we obtain that for sufficiently early times $D\gg\Pi_P,\Pi_H$, yielding an early regime in which both elasticity and compressibility contribute to the peeling process. However, for injection of mass at small but finite time-scales there is an initial value of the characteristic gauge pressure (denoted hereafter by $p_0^*$),  scaling by which yields $D\sim O(1)$ for early times. This sets the requirement $\Pi_P,\Pi_H\ll1$ as a condition for the appearance of the early time regime  (i.e. large gauge pressure compared with background pressure and large displacement to prewetting thickness  ratio). In this early time regime the leading order of equation (\ref{GE_sbs_H}) is a porous-medium-equation of order $2.5$ for the variable $D^2$. Applying ZKB's solution \citep{Barenblatt1952PME} yields the time propagation rate $X_F=O(\mathfrak{T}^{2/7})$ and the peeling dynamics are given by,
\begin{subequations}\label{fundamental_early}
\begin{equation}
D(X,\mathfrak{T}) = {\left( {\frac{1}{5}} \right)^{ - 1/7}}{\mathfrak{T}^{ - 1/7}}\left[ {{C_1} - \frac{3}{{35}}{{\left( {\frac{1}{5}} \right)}^{ - 4/7}}{X^2}{\mathfrak{T}^{ - 4/7}}} \right]_ + ^{1/3}{\kern 1pt} ,
\end{equation}
\begin{equation}
{C_1} = {\left( {\frac{{2M}}{{{S_1}}}} \right)^{6/7}},\,\,\,\,{S_1} = \sqrt {\frac{{35\pi }}{3}} \frac{{B(1/2,5/3)}}{{\Gamma (1/2)}}\,,
\end{equation}
\end{subequations}
where $B$ is the beta function, $\Gamma$ the gamma function and $(s)_{_+}=\max(s,0)$. 
As the added mass propagates and expands into the substrate, the gas pressure decreases and thus $D$ decreases, eventually invalidating the requirement of $D\gg\Pi_P,\Pi_H$. This sets a validity time range of  $\mathfrak{T}\ll 5 C_1^{7/3}/ \text{max}(\Pi_P^7,\Pi_H^7)$ for (\ref{fundamental_early}) based on $D(X=0)$.

For the case of   $\Pi_H/\Pi_P=h_0/(p_a/k)\gg 1$ (corresponding to negligible effects of elasticity and dominant effects of low-Mach-number gas compressibility) an intermediate regime exists where $\Pi_P\ll D \ll \Pi_H$.
In this regime, the leading order of equation (\ref{GE_sbs_H}) is a porous-medium-equation of order 2 for $D$. Thus the solution will transition to a propagation rate of $X_F=O(\mathfrak{T}^{1/3})$  and the resulting profile,
\begin{subequations}\label{rigid_limit}
\begin{equation}
D(X,\mathfrak{T}) = {\left( {\frac{{\Pi_H^2}}{{2}}} \right)^{ - 1/3}}{\mathfrak{T}^{ - 1/3}}{\left[ {C_2 - \frac{1}{{12}}{{\left( {\frac{{\Pi_H^2}}{{2}}} \right)}^{ - 2/3}}{X^2}{\mathfrak{T}^{ - 2/3}}} \right]_ + }\,,
\end{equation}
\begin{equation}
{C_2} = {\left( {\frac{{2M}}{{{\Pi _H}{S_2}}}} \right)^{2/3}},\,\,\,\,{S_2} = \sqrt {12\pi } \frac{{B(1/2,2)}}{{\Gamma (1/2)}}\,,
\end{equation}
\end{subequations}
will emerge in intermediate times with a validity range of $ 2C_2^{3} / \Pi_H^5 \ll \mathfrak{T} \ll 2 C_2^{3} / \Pi_H^3 \Pi_P^2$. Solution (\ref{rigid_limit}) represents the limit of dominant gas compressibility, and is identical to the evolution of compressible low-Reynolds-number gas flow in rigid configurations. The relevant time-scale of this limit is $t^*=\mu/p^* \varepsilon^2$.

Alternatively, for $\Pi_P / \Pi_H=p_a/kh_0 \gg 1$, a different intermediate region exists for which $\Pi_H\ll D\ll \Pi_P$ and the leading order evolution equation is a porous-medium-equation of order 4 for $D$, yielding

\begin{subequations}\label{incomp_limit}
\begin{equation}
D(X,\mathfrak{T}) = {\left( {\frac{1}{{4}}} \right)^{ - 1/5}}{\mathfrak{T}^{ - 1/5}}\left[ {C_3 - \frac{3}{{40}}{{\left( {\frac{1}{{4}}} \right)}^{ - 2/5}}{X^2}{\mathfrak{T}^{ - 2/5}}} \right]_ + ^{1/3}\,,
\end{equation}
\begin{equation}
{C_3} = {\left( {\frac{{2M}}{{{\Pi _P}{S_3}}}} \right)^{6/5}},\,\,\,\,{S_3} = \sqrt {\frac{{40\pi }}{3}} \frac{{B(1/2,4/3)}}{{\Gamma (1/2)}}\,,
\end{equation}
\end{subequations}
with an $X_F=O(\mathfrak{T}^{1/5})$ spread-rate, typical of the early time propagation of the incompressible peeling problem \citep[e.g.][]{elbaz2016axial}, and a validity range of  $ 4C_3^{5/3} / \Pi_P^5 \ll \mathfrak{T} \ll 4 C_3^{5/3} / \Pi_H^5$.

By setting larger values for $\Pi_H$ or $\Pi_P$, propagation dynamics may skip or move across a certain stage in the sequence. All solutions will ultimately settle on $X_F=O(\mathfrak{T}^{1/2})$ propagation as $\mathfrak{T}\to \infty$ whether be it the prewetting thickness ratio $\Pi_H$ or background to gauge pressure ratio $\Pi_P$ the final regularization mechanism which linearizes (\ref{GE_sbs_H}). 

The evolution of the solution through the various regimes and corresponding propagation rates is validated numerically in figure 3(a).  A flow-chart illustrating the transitions and presenting the requirements for the different limits, as well as the time-ranges in which the limits are valid, is presented in figure 3(b). The validity range for each limit is calculated by requiring the appropriate order of magnitude of $D$ for the examined limit from the solutions (\ref{fundamental_early}), (\ref{rigid_limit}) and (\ref{incomp_limit}) at $X=0$. The conditions $(1)-(6)$ presented in figure 3(b) may be represented in dimensionless form as $(1)$ $\Pi_P \gg 1$ and $\Pi_H\ll 1$; $(2)$ $\Pi_P \ll 1$ and $\Pi_H\ll 1$; $(3)$ $\Pi_P \ll 1$ and $\Pi_H\gg 1$; $(4)$ $\Pi_H\ll \Pi_P$; $(5)$ $\Pi_H\sim \Pi_P$; and $(6)$ $\Pi_H\gg \Pi_P$. An additional solution with velocity-slip (dashed line) was also considered along with its no-slip counterpart marked by $(2)+(5)$. For intermediate times Knudsen-diffusion is shown to mildly alter the spread rate, but this effect is reduced at the early time limit  (\ref{fundamental_early}) as well as in the linearized regime.  Figure 3(a) is supplemented by figure 4 which presents the numerical deformation profiles for various limits and the convergence of the numerical profile to the theoretical results presented in (\ref{fundamental_early}), (\ref{rigid_limit}) and (\ref{incomp_limit}), corresponding to panels (a), (b) and (c), respectively. 




\begin{figure}
\centering
\includegraphics[width=0.85\textwidth]{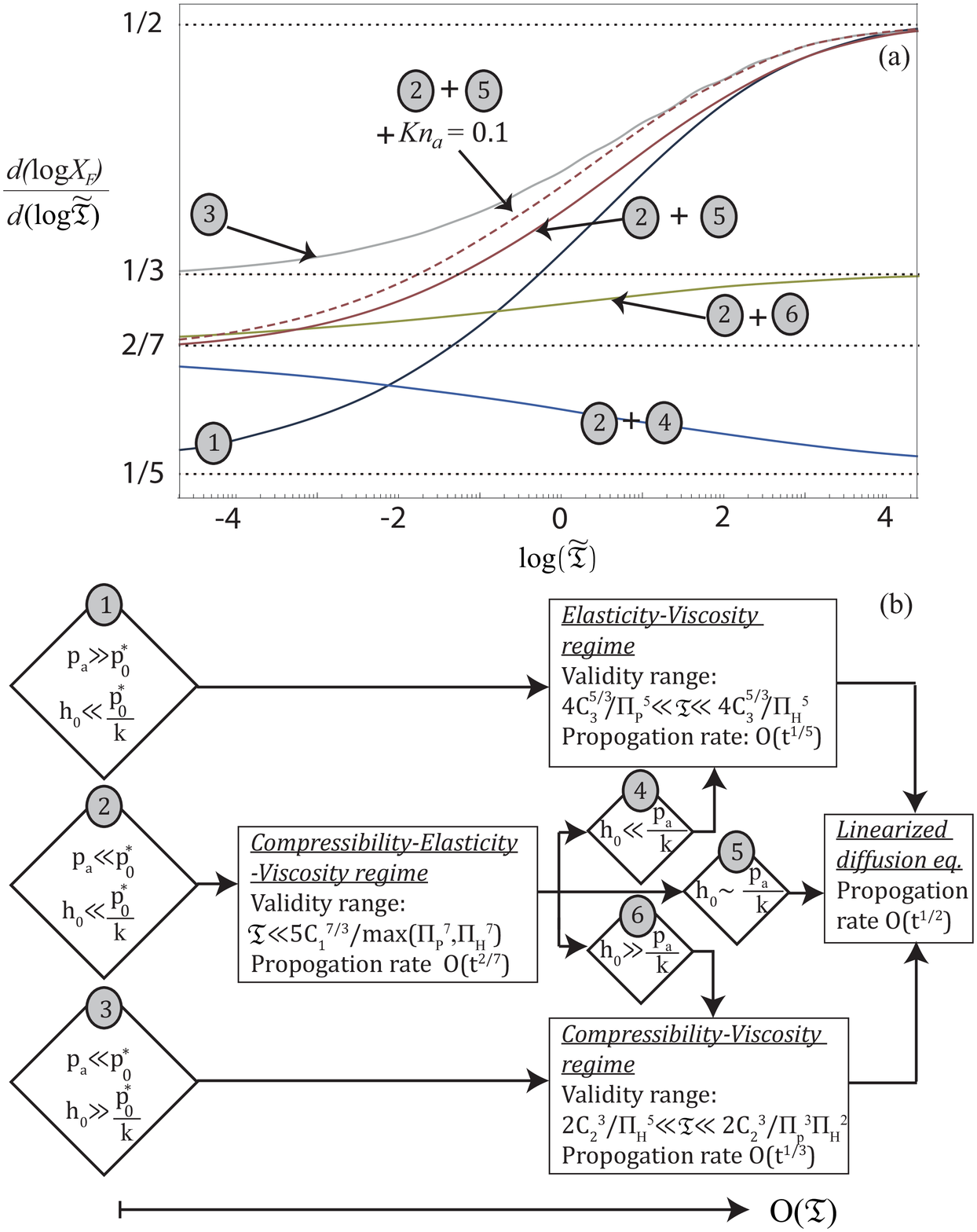}
\caption{Propagation regimes of gaseous viscous peeling in a 2D gap bounded by linearly elastic substrates for different time-scales and physical parameters. (a) Numerical computation of the propagation front velocity (logarithmic derivative) from (\ref{GE_sbs_H}) vs. the theoretical limits given by (\ref{fundamental_early})-(\ref{incomp_limit}).   Line $(1)$ corresponds to $(M,\Pi_H,\Pi_P)=(3.8 \times10^{-3},2\times 10^{-3},2)$; line $(2)+(4)$ corresponds to $(M,\Pi_H,\Pi_P)=(2.86\times 10^{-4},0,1\times 10{-2})$;  line $(2)+(5)$ corresponds to $(M,\Pi_H,\Pi_P)=(2.95 \times10^{-4},5\times 10^{-3},1\times 10^{-2})$; line $(2)+(6)$ corresponds to $(M,\Pi_H,\Pi_P)=(2.77 \times10^{-4},5\times 10^{-3},0)$; line $(3)$ corresponds to $(M,\Pi_H,\Pi_P)=(5.35 \times10^{-4},0.15,10^{-4})$.  Time is rescaled according to the average time presented in the figure, $\tilde{\mathfrak{T}}=10^7\mathfrak{T}$. (b) A flow-chart diagram describing the different dominant balances, transitions between regimes and propagation time-scales as a function of initial gauge pressure $p_0^*$, background pressure $p_a$, channel stiffness $k$ and prewetting layer thickness $h_0$.}
\label{figure3}
\end{figure}

\begin{figure}
\centering
\includegraphics[width=1\textwidth]{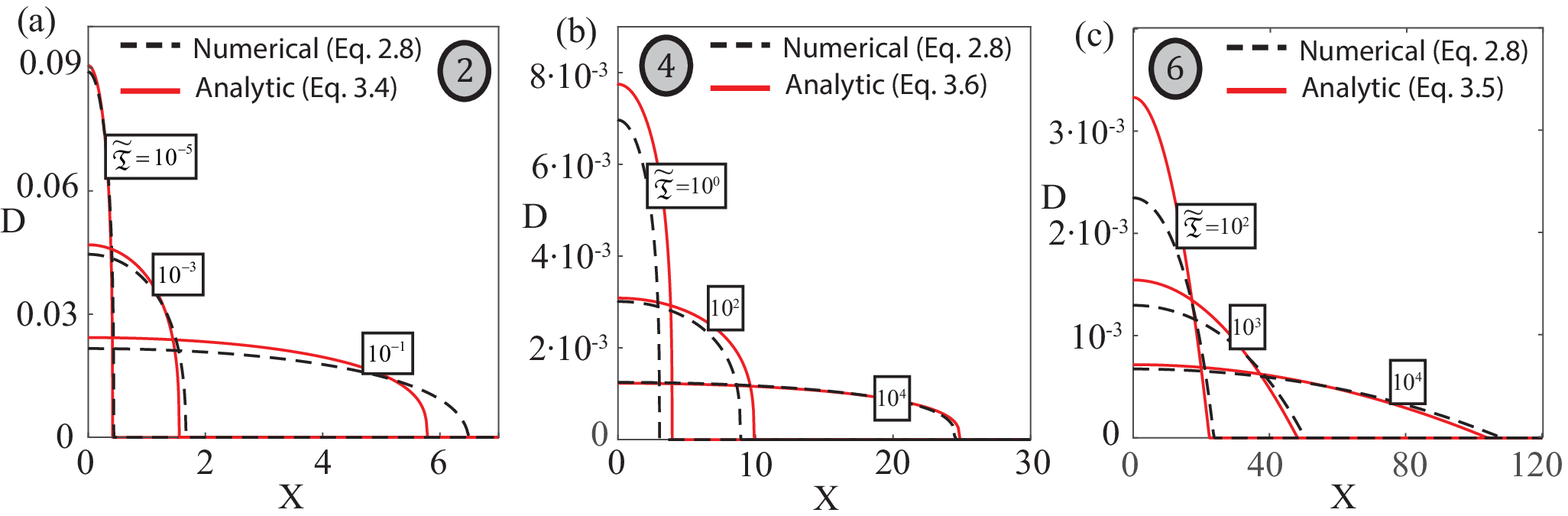}
\caption{Convergence of the numerical deformation profiles to the theoretical results (\ref{fundamental_early}), (\ref{rigid_limit}) and (\ref{incomp_limit}) corresponding to the respective values of ($M,\Pi_H,\Pi_P$) presented in figure \ref{figure3}. Panels (a), (b) and (c) present convergence along lines (2)+(6) (stage (2)), (2)+(4) (stage (4)) and (2)+(6) (stage (6)), respectively. }
\label{figure4}
\end{figure}

\subsection{Self-similar solution with weak rarefaction effects for $\Pi_P\rightarrow\Pi_H$}
For the limit of $\Pi_H\rightarrow\Pi_P$ (or $k\rightarrow p_a/h_0$), applying $f=D+\Pi_H=D+\Pi_P$ allows to obtain an additional self-similar solution for the case of suddenly applied fixed inlet pressure. In this limit both the viscous and Knudsen diffusion terms of equation (\ref{GE_sbs_H}) will enforce a $O(\mathfrak{T}^{1/2})$ spread-rate and an exact self-similar solution with velocity-slip may be attained for $f(\eta)$, where $\eta=X\mathfrak{T}^{-1/2}$. 

Substitution of $f(\eta)$ into (\ref{GE_sbs_H}) yields the self-similar boundary value problem,
\begin{subequations}\label{self_similar_BVP}
\begin{equation}
{f^5}'' + 10 \sigma Kn_a \Pi_H^2 {f^3}'' + \frac{{5\eta }}{2}{f^2}' = 0\,,
\end{equation}
supplemented by
\begin{equation}
f(0) = 1 + {\Pi _H},\,\,\,\,\,f(\infty ) = {\Pi _H}\,,
\end{equation}
and
\begin{equation}
\int_0^\infty  {{H^2}(X,\mathfrak{T})dX = M} {\mathfrak{T}^{1/2}},\,\,\,M = \int_0^\infty  {{f^2}(\eta )d\eta }\,.
\end{equation}
\end{subequations}

\begin{figure}
\centering
\includegraphics[width=0.9\textwidth]{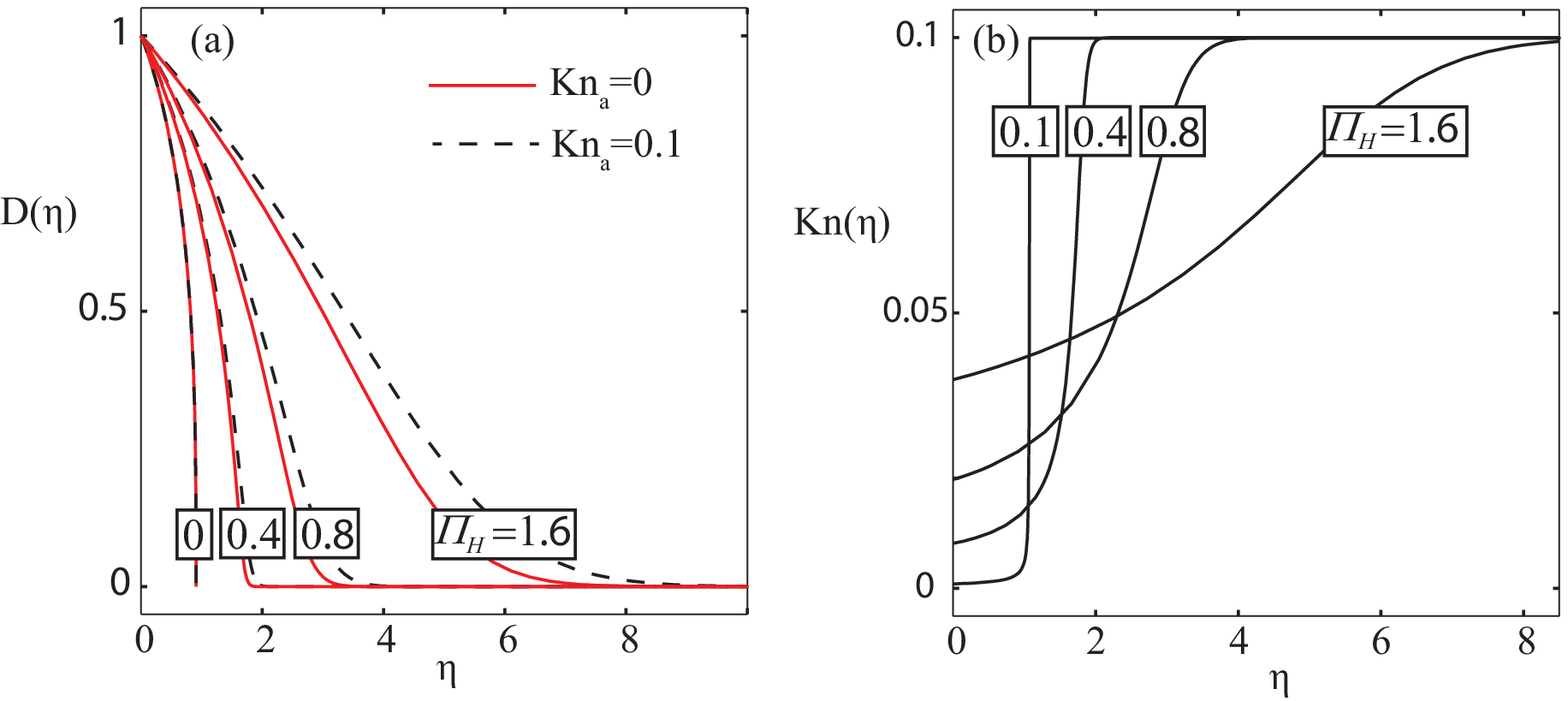}
\caption{Self-similar solutions with weak-rarefaction effects for $\Pi_P=\Pi_H$  and $\sigma=1$. (a) Self-similar deformation profile as a function of the prewetting thickness ratio $\Pi_H$ for $Kn_a=0$ (smooth lines) and  $Kn_a=0$ (dashed lines). (b) Local Knudsen as a function of $\eta$ for various values of $\Pi_H$ for $Kn_a=0.1$. Solution based on numerical integration of (\ref{self_similar_BVP}).}
\label{figure5}
\end{figure}

Self-similar profiles for various values of $\Pi_H$ are presented in figure \ref{figure5}(a) for $ Kn_a=0$ (smooth lines) and $\sigma=1,\, Kn_a=0.1$ (dashed lines). The effect of weak rarefaction is shown to increase the speed of gas propagation, and reduce the gradients of the deformation. This effect decreases as $\Pi_H$ and $\Pi_P$ decrease, since $Kn_a$ is defined ahead of the front, while the local Knudsen decreases as the gap and pressure increase (see (\ref{Pis})). For small $\Pi_P$ and $\Pi_H$, the pressure and gap in the peeled region are greater compared with the background pressure $p_a$ and prewetting layer thickness $h_0$. This yields a significantly smaller effective Knudsen number in the peeled region, as illustrated in figure \ref{figure5}(b). 


\section{Concluding remarks}
The propogation of a gas into micron-sized configurations with linearly elastic boundaries is governed by interaction between effects of low-Mach-compressibility, weak rarefaction, elasticity and viscosity. While exact solutions of the governing nonlinear evolution equation are not available, several limiting cases allow solution by self-similarity. These limits correspond to different physical regimes, including: (i) dominant balance between compressiblity and viscosity, (PME of order $2$) characterizing compressible flow in rigid micro-channels, (ii) dominant balance between elasticity and viscosity, (PME of order $4$) characterizing incompressible flow in elastic micro-channels, and (iii) dominant balance involving viscosity, elasticity and compressiblity (PME of order $2.5$). During gas film propagation, the flow-field transitions between the aforementioned regimes and corresponding exact solutions. A map of these transitions was  presented as a function of the prewetting layer thickness, the background pressure and stiffness of the spring array. The case where $k h_0\approx p_a$ represents symmetry between compressibility and elasticity, and allowed to obtain an additional self-similar solution accounting for weak rarefaction effects. 

While the steady-state solution presented in \S 3.1 is implicit, explicit solutions may be obtained for the same physical limits examined in the transient dynamics, as presented in \S 3.2. For negligible slip $Kn_0\rightarrow0$, steady-state solutions corresponding to the Barrenblatt self-similar limits (\ref{fundamental_early})-(\ref{incomp_limit}) can be presented by the relation
\begin{equation}\label{steady_sol_sbs_exp}
D(X)=\left[\left( (D(1)+C)^{N}-(D(0)+C)^{N} \right)X+(D(0)+C)^{N}\right]^{\frac{1}{N}}-C,
\end{equation}
where $(N,C)=(5,\Pi_P)$ for the limit of $\Pi_P,\Pi_H\ll D$ (viscous-elastic-compressibility regime) or the symmetric case of $\Pi_P=\Pi_H$;  $(N,C)=(2,\Pi_P)$ for the limit of $\Pi_H\gg \Pi_P,D$ (viscous-compressibility regime); $(N,C)=(4,\Pi_H)$ for the limit of $\Pi_P\gg \Pi_H,D$ (viscous-elastic regime) and finally, $(N,C)=(1,0)$ for the linear limit of $\Pi_P,\Pi_H\gg D$ (viscous regime) .

\bibliographystyle{jfm}
\bibliography{Bib_File}
\end{document}